\documentclass[prl,aps,nofootinbib,showkeys,showpacs,twocolumn]{revtex4} 

\usepackage{epsfig}
\usepackage{graphicx}% Include figure files
\begin{document}
\date{\today}

%\title{Density Matrix Functional Theory for Pairing}
\title{Density-matrix functionals for pairing in mesoscopic superconductors}

\author{Denis Lacroix} \email{lacroix@ganil.fr}
\affiliation{GANIL, CEA and IN2P3, Bo\^ite Postale 55027, 14076 Caen Cedex, France}
\author{Guillaume Hupin} \email{hupin@ganil.fr}
\affiliation{GANIL, CEA and IN2P3, Bo\^ite Postale 55027, 14076 Caen Cedex, France}

\begin{abstract}
A functional theory based on single-particle occupation numbers is developed for pairing. This functional, that generalizes the 
BCS approach, directly incorporates corrections due to particle number conservation. The functional is benchmarked  
with the pairing Hamiltonian and reproduces perfectly the energy for any particle number and coupling.
\end{abstract}

\pacs{4.20.-z 71.15.Mb, 21.60.Fw}
\keywords{pairing, functional theory, algebraic models.} 

\maketitle

While  the Bardeen-Cooper-Schrieffer (BCS) microscopic
theory \cite{Bar57} provides a suitable description of superconductivity 
in the macroscopic limit, it is not accurate enough for 
mesoscopic systems, such as nuclei, atomic clusters, quantum
dots, fullerenes, nanotubes, or ultrasmall metallic grains 
\cite{Rin80,Von01,Bri05,Leg07}. Standard BCS approach to superconductivity 
has some drawbacks: (i) non negligible corrections due to finite size effects 
are necessary in mesoscopic systems; (ii)
in condensed-matter and/or 
nuclear physics, BCS theory starting from the bare many-body 
interaction cannot be considered either as a numerically tractable approach nor as a predictive 
theory. To overcome difficulty (ii), specific functional theories based not only on the local density 
$\rho(r)$ but also on the anomalous density $\kappa(r,r')$ are used \cite{Lud05,Ben03}. Guided by the Hamiltonian 
case, the energy is decomposed as:
\begin{eqnarray}
\label{eq:e00}
{\cal E} [\rho, \kappa ] & = & {\cal E} [\rho] + {\cal E}_{\rm Pair} [\kappa, \kappa^*] , \label{eq:func}
\end{eqnarray} 
where ${\cal E} [\rho]$ is often taken as the energy density functional without pairing, while  ${\cal E}_{\rm Pair} [\kappa, \kappa^*]$
is the extra contribution due to pairing.  Eq. (\ref{eq:func}) turns out to be 
very accurate to deal with (ii) for instance in nuclear physics. 
In that case, systems with 10-200 constituents are considered and  additional finite size corrections are necessary \cite{Rin80,Bla86,Bri05}. 
However, recent studies have shown that techniques generally used to restore good particle number should be handled with care when combined with density functional theories \cite{Dob07,Lac09,Ben09,Dug10,Rob10}. In particular, 
unless new methods able to properly  treat finite size effects and more generally configuration mixing 
within functional theory, most of the functional designed during the last 30 years have to be revisited\cite{Dug09}.    
These difficulties question the possibility to use symmetry breaking within a functional theory.

The goal of the present work is to provide a new theoretical framework for pairing in finite systems that avoids difficulties 
recently encountered in functional theories and that can be applied easily be implemented in current functionals. 
Here, we follow the idea of 
Gilbert\cite{Gil75,Pap07,Lac09b} and  seek for a functional of occupation probabilities $n_i$ and
natural orbitals  $\varphi_i$, i.e. ${\cal E} \equiv  {\cal E} [n_i, \varphi_i ]$. Note in passing, that 
(\ref{eq:func}) already enters into the class of Gilbert functionals.  Indeed, the pairing energy
is generally written as ${\cal E}_{\rm Pair} [\kappa, \kappa^*] =  \sum_{i j} \bar v^{\kappa \kappa}_{i j k l} \kappa_{ij} 
\kappa^*_{kl}$ where $\bar{v}^{\kappa \kappa}$ denotes the effective two-body kernels  in the 
particle-particle and hole-hole channels.  In practice, both  $\rho$ and $\kappa$ are computed using a quasi-particle (QP) vacuum trial state that can be seen as a generalization of the Kohn-Sham Slater determinant.
If the energy is minimized in the canonical basis, the 
trial state, denoted by $| \phi \rangle$, takes a BCS like form  $
%\begin{eqnarray}
| \phi [n_i, \varphi_i ] \rangle   =  \prod_{i} \left( 1 + x_i a^\dagger_i a^\dagger_{\bar i} \right) | 0 \rangle$
%\label{eq:bcsstate}
%\end{eqnarray}
with $x_i = \sqrt{n_i/(1-n_i)}$ and 
where $| 0 \rangle$ corresponds to the particle vacuum while 
$\{ a^\dagger_i , a^\dagger_{\bar i} \}$ are associated to doubly degenerated canonical 
 states $\{\varphi_i , \varphi_{\bar i}\}$ with
 occupation probabilities $n_i$. In the canonical basis, $\kappa$ becomes block diagonal with 
 $\kappa_{i\bar i}  = \sqrt{n_i (1-n_i)}$ and finally leads to  an energy  functional of the BCS occupation 
 numbers.
%\begin{eqnarray}
%{\cal E}_{\rm Pair} [n_i, \varphi_i] & = &  \sum_{i j} \bar v^{\kappa \kappa}_{i \bar i j \bar j} 
%\sqrt{n_i (1-n_i)}  ~ \sqrt{n_j (1-n_j)}. \label{eq:bcsfunc}
%\end{eqnarray}  
Functional based on BCS suffers for instance from the absence of pairing at weak coupling, it also misses 
part of the pairing effects at strong coupling (see for instance \cite{Pap07}). These defects can be cured 
by considering many-body trial states projected onto good particle number. In that case, a reference QP state is first 
introduced, onto which the projection is made. The resulting energy becomes a rather complicated 
functional of the reference state \cite{She00}. 
However, 
serious difficulties appear when projection technique made before or after variation is combined with functional theories 
\cite{Dob07,Lac09,Ben09,Dug09,Rob10}. 
 
Here, we use a completely different strategy and consider the projected state directly as the trial state.  This state, called 
hereafter, Projected BCS state is written in its canonical basis as:
\begin{eqnarray}
| N \rangle & \equiv &  \frac{1}{\sqrt{N!}}(\Gamma^\dagger)^N | - \rangle, \label{eq:pbcsstate}
\end{eqnarray}
where $\Gamma^\dagger = \sum_i x_i b^\dagger_i$ with $b^\dagger_i = a^\dagger_i a^\dagger_{\bar i}$. $N$ 
denotes here the number of pairs. From this state, we define the occupation number and  two-body correlation matrix elements
through:
\begin{eqnarray}
n_i &\equiv& \frac{\langle N  |a^\dagger_i a_i | N \rangle }{\langle N | N \rangle }, ~~  
R_{ij} \equiv \frac{\langle N |b^\dagger_i b_j | N \rangle }{\langle N | N \rangle }. \label{eq:defnc}
\end{eqnarray}

Guided by the Hamiltonian framework, we propose to generalize the pairing energy and writes
\begin{eqnarray}
{\cal E}_{\rm Pair} [R, \varphi_i] & = &  \sum_{i j} \bar v^{\kappa \kappa}_{i \bar i j \bar j} R_{ij}. \label{eq:genfunc}
\end{eqnarray}  
where $v^{\kappa \kappa}$ can eventually depend on the density of the projected state. By doing so, difficulties
observed with projection are avoided.
The main result of the present paper is to show that the correlation can be accurately written as a functional of occupation 
numbers of the projected state and account for finite size correction.  
  
According to the definitions (\ref{eq:defnc}), both $n_i$ and $R_{ij}$ can be written as functionals of the 
parameter set   $\{ x_i \}$:
\begin{eqnarray}
\left\{
\begin{array} {lll}
n_i &=& \displaystyle  N |x_i|^2 \frac{I_{N-1}(i)}{I_N},   \\
R_{ij} &=&  \displaystyle N x^*_i x_j \frac{I_{N-1}(i,j)}{I_N} ~~{\rm for} ~~ (i \neq j),  \\
\end{array}
\right. 
\label{eq:nicipbcs}
\end{eqnarray}
while for $i = j$, $R_{ij} = \displaystyle n_i$.  In these expressions, the following quantities have been defined (for any K such that $1 < K \leq N$ and any $n$ with $n < \Omega$): 
%\footnote{Note that in the following, we will not make the distinction between
% $I_K$ and the other parameters, using the convention that $I_K() = I_K$}
\begin{eqnarray}
%\left\{
%\begin {array} {lll}
%I_K &=& \displaystyle  \sum^{\neq}_{(j_1, \cdots j_{K})} |x_{j_1}|^2 \cdots |x_{j_{K}}|^2 \\
I_K(i_1,  \cdots , i_n ) &=& \displaystyle 
\sum^{\neq}_{(j_1, \cdots j_{K}) \neq (i_1,  \cdots , i_n)} |x_{j_1}|^2 \cdots |x_{j_{K}}|^2, 
%\end{array}
%\right.
 \label{eq:ikijdef}
\end{eqnarray}
where $\sum^{\neq}$ means that the summation is made over all $(j_1, \cdots j_{K})$ different from each other while 
$(j_1, \cdots j_{K}) \neq (i_1,  \cdots , i_n)$ adds the constraint that all $j_n$ are different from $(i_1,  \cdots , i_n)$. 
%Expressions (\ref{eq:nicipbcs}) can  directly be obtained, for instance, by expanding (\ref{eq:pbcsstate}).
{From the above discussion and as can be intuitively from the expression of the trial state (\ref{eq:pbcsstate}), 
the energy can be written as an explicit functional of the $\{x_i\}$.
Unfortunately, the complexity of this functional prohibits its use\footnote{Note that recently, it has been shown that such a minimization could eventually be performed numerically using specific recurrence relations satisfied by the $I_K$\cite{Van86,San08,Lac10}.}. A second difficulty of using 
$\{x_i\}$ as variational parameters is that they are not easily connected to quantities like occupation numbers 
adding complexity in physical interpretation. The BCS example illustrates that the energy 
can directly be written as a functional of the $\{n_i\}$ but is insufficient to precisely grasp the physics of 
pairing in finite systems.}   
%\begin{eqnarray}
%\left\{
%\begin {array} {lll}
%I_K(i_1,  \cdots , i_n ) &=& \displaystyle I_K(i_1,  \cdots , i_n , k) \\
%&+& \displaystyle K  |x_k|^2 I_{K-1}(i_1,  \cdots , i_n, k) \\
%\\
%I_K(i_1,  \cdots , i_n) &=&  \sum_{ k\neq (i_1, \cdots i_{K})}  |x_k|^2 I_{K-1} (i_1,  \cdots , i_n),
%\end{array}
%\right.
% \label{eq:ikij}
%\end{eqnarray}
%where the first equation is valid  for all    $k \neq (i_1,  \cdots , i_n ) $. 

Despite the complex relations between the occupation 
numbers and the $\{ x_i \}$ (Eq. (\ref{eq:nicipbcs})), it is shown 
below that (i) the $\{ x_i \}$ can be accurately replaced by a functional of the $\{ n_i \}$; (ii) using this functional, 
the energy itself becomes a functional of the occupation probability only, 
that provides a very accurate description of the pairing Hamiltonian energy.  
Different relations given below are directly derived from recurrence relations existing for the $I_K$\cite{Van86,San08,Lac10}  after tedious but straightforward calculations. We only give here main results are summarized here while technical details will be given elsewhere \cite{Lac10}.  

%By introducing the notation 
%\begin{eqnarray}
%\alpha_K  (i_1, \cdots, i_n ) &\equiv& \frac{I_K (i_1, \cdots, i_n) }{K I_{K-1}(i_1, \cdots, i_n)},
%\end{eqnarray}
Eqs. (\ref{eq:nicipbcs}) can be rewritten as
\begin{eqnarray}
n_i &=& \frac{|x_i|^2}{|x_i|^2 +  \alpha_i},~
%  \label{eq:nipbcs}  
% \\
R_{ij} = \frac{x^*_i x_j} {|x_i|^2 - |x_j|^2}  (n_i - n_j ), \label{eq:nicijpbcs2}
\end{eqnarray} 
where $\alpha_i \equiv I_N(i)/(NI_{N-1}(i))$.

An interesting connection with the BCS theory can be made by inverting the first expression in (\ref{eq:nicijpbcs2}) as 
\begin{eqnarray}
|x_i|^2 &=& \left(\frac{n_i}{ 1 -n_i} \right)  \alpha_i.
\end{eqnarray}
Reporting in the second equation of (\ref{eq:nicijpbcs2}), leads to
\begin{eqnarray}
R_{ij}  &=& \sqrt{n_i (1-n_i) n_j (1-n_j) \alpha_i \alpha_j}  \nonumber \\
&& \times \frac{n_i -n_j}{ n_i (1-n_j) \alpha_i - n_j (1-n_i) \alpha_j }. \label{eq:cijpbcs_alpha}
\end{eqnarray}
Assuming $\alpha_i = \alpha_j=1$, this expression identifies with the BCS limit. 
Therefore, the physics associated with particle number restoration is all contained in the set of $\{ \alpha_i \}$ parameters.
%It could indeed be shown that any of the $\alpha_K$ is equal to $1$ when the PBCS wave-function is replaced 
%by the BCS one. 
%Therefore, the physics associated with particle number restoration is all contained 
%in the set of $\{ \alpha_i \}$ parameters. To obtain an explicit form of the functional, using equations (\ref{eq:ikij}), 
%it is first useful to realize that $\alpha_N$ is connected to $\alpha_1$ by a hierarchy of equations 
%between the $\alpha_K$ and the $\alpha_{K-1}$: 
% \begin{eqnarray}
%\alpha_{K}(i_1, \cdots, i_n ) &=&   \frac{1}{K} \sum_{j }^{\neq}  |x_j|^2  \frac{\alpha_{K-1} (i_1, \cdots, i_n ,j) }{|x_j|^2 + \alpha_{K-1} 
%(i_1, \cdots, i_n,j)},  \nonumber
%\end{eqnarray}
%valid for $K > 1$. 
The $\alpha$ parameters can be written as a function of the occupation numbers 
by performing a 
%This hierarchy can be used to obtain a systematic 
$(1/N)$ expansion %of the $\alpha$ parameters 
assuming that
%Assuming that 
the leading order identifies with the BCS limit 
%and that the expansion is independent of $K$, 
one 
obtains\footnote{Note that, additional terms tested numerically as negligible and appearing 
at the second order approximation are omitted here.} \cite{Lac10}:
\begin{eqnarray}
\alpha_i &=& 1 - \frac{1}{N} n_i \nonumber \\
&+& \frac{1}{N(N-1)}  \sum_{j \neq i} n_j^2 [ 1  -  (n_i + n_j)  ] \nonumber \\
&+& \frac{1}{N(N-1)(N-2)}  \sum_{(k,j) \neq i}^{\neq} n_j^2 n^2_k  \left[2 -  (n_i  + n_j +n_k) \right]  \nonumber \\
&+&\cdots \label{eq:explicit1}
\end{eqnarray}  
which is nothing but $\alpha_i$ written as an explicit functional of the occupation numbers. 
{By replacing $\alpha_i$ given above in $R$ and then the expression of $R$ in (\ref{eq:genfunc}), the energy 
become an explicit functional of the occupation numbers through the sequence
\begin{eqnarray}
{\cal E} [\rho, R, \varphi_i] & \rightarrow & {\cal E} [\{x_i\}, \varphi_i]  \rightarrow  {\cal E} [\{n_i\}, \varphi_i]. \label{eq:genfunc2}
\end{eqnarray}   
%This functional include finite size corrections. 
In practice, the new expression 
we obtained for ${\cal E}$ can directly replace the BCS expression in theories either based on a Hamiltonian 
or directly formulated in a density functional framework. Since the functional is directly formulated in terms 
of natural orbitals and occupancies, one should add specific constraint during the minimization.  
Below, the quantity:
\begin{eqnarray}
{\cal E}' [\{n_i\}, \varphi_i] & = & {\cal E} [\{n_i\}, \varphi_i] - \mu \left( \sum n_i -N \right), \label{eq:genfunc2}
\end{eqnarray}   
is directly minimized.}
% where the Lagrange multiplier $\mu$ is added to insure particle number conservation.  
\begin{figure}[htbq]
%\begin{center}
\includegraphics[width = 6.cm]{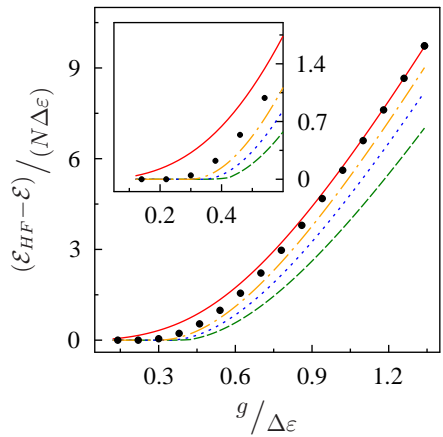}
%\end{center}
\caption{ \label{fig2:pbcs} (Color online) Difference between the Hartree-Fock energy ${\cal E}_{HF}$ and the exact Richardson solution (red solid line) of ref. \cite{San08}, generally referred as the condensation energy, compared to result obtained with successive approximation using expression (\ref{eq:explicit1}) for 8 particles. The
dashed, dotted, dot-dashed lines and filled circles  
correspond respectively to the BCS, first, second and third order correction. In insert, a focus on the 
weak coupling region is shown. Note that the PBCS energy (not shown here) cannot be distinguished from the exact result. }
\end{figure}

%--------HERE Richardson----------

In order to test the functionals developed here, we consider an even system of A particles interacting through the "picket" fence Hamiltonian of the form \cite{Ric64} 
\begin{eqnarray}
H = \sum_{i} \varepsilon_i (a^\dagger_i a_i + a^\dagger_{\bar i} a_{\bar i}) - g \sum_{i,j }  a^\dagger_i 
a^\dagger_{\bar i} a_{\bar j} a_{j},
\end{eqnarray} 
with doubly degenerate equidistant levels and with a level spacing $\Delta \varepsilon$.
In the following, $N=A/2$ denotes the number of pairs while $\Omega = 4N$ is the single-particle space size.
For this Hamiltonian, the energy  simply writes $
{\cal E} (n_i , R_{ij}) \equiv  2 \sum_i  \varepsilon_i n_i - g  \sum_{ij} R_{ij}$. Note that, this functional 
interaction corresponds to (\ref{eq:genfunc}) with $v^{\rho \rho}_{i \bar i j \bar j}=v^{\kappa \kappa}_{i \bar i j \bar j} = g$.
The pairing Hamiltonian can be solved exactly and therefore is particularly suitable for benchmarking 
approximation for pairing correlation\cite{Von01,Duk04}.

%--------HERE Richardson----------

An illustration of successive corrections beyond the BCS approximation is given in figure \ref{fig2:pbcs}. 
{Results are obtained using the functional form of $R$ where $\alpha_i$ are replaced by Eq. (\ref{eq:explicit1}) trunctated 
at a given order. Then, Eq. (\ref{eq:genfunc2}) is directly minimized starting from the BCS solution and making variation 
of occupation numbers between 0 and 1. A quadratic programming (QP) method is used for the minimization leading to 
rapid convergence tested up to 400 particles.}
Systematic improvement is observed as higher orders in the correction are included. The above functional has however two major drawbacks.
First, it is rather complicated to use. Second, while few terms are necessary to get a perfect result in the strong coupling, the convergence 
is rather slow in the weak coupling regime. It could indeed be shown that all terms in the expansion up to order $1/N!$ contribute
equally in the Hartree-Fock limit. This directly stems from the inadequacy of the BCS functional 
in the weak coupling limit. For instance, a direct use of perturbation theory leads to much better results underlying 
the role of $2$ particles-$2$ holes excitations (see for instance discussion in \cite{San08}).
\begin{figure}[htbq]
\includegraphics[width = 7.cm]{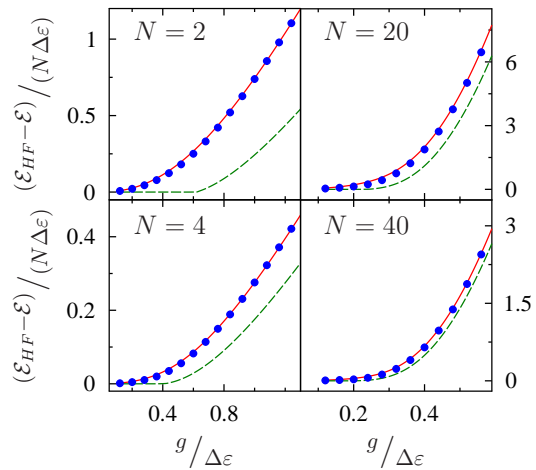}
\caption{ \label{fig3:pbcs} (Color online) Comparison between the exact condensation energy (solid line), the BCS (dashed line) 
and the result obtained by 
minimizing the density matrix functional (filled circles) with the approximation (\ref{eq:a0a1}) for various particle numbers and coupling strength . 
Again in all cases, the PBCS result is in perfect agreement with the exact one.}
\end{figure}
This difficulty could only be overcome by considering these terms explicitly, which in the form given by (\ref{eq:explicit1}) becomes impractical for numerical implementation as $N$ increases. A simple linear expression, i.e.  
\begin{eqnarray}
\alpha_i = a_0 + a_1 n_i , \label{eq:linear}
\end{eqnarray}
can however be found by considering all terms in the expansion and approximating sums in (\ref{eq:explicit1}) according to:
\begin{eqnarray}
&&\frac{1}{N(N-1)}  \sum_{j \neq i}  \rightarrow   
\frac{1}{N^2}  \sum_{j} ,  \nonumber \\
&&\frac{1}{N(N-1)(N-2)}  \sum_{(k,j) \neq i}^{\neq}   \rightarrow \frac{1}{N^3}  \sum_{j k}  , \cdots \nonumber
\end{eqnarray} 
A straightforward calculation then gives:
\begin{eqnarray}
a_1 &=& -  \frac{1}{N}  \frac{1 - s_2^N}{1 - s_2} , ~~a_0  =  1  - (s_2-s_3)  \frac{\partial a_1}{\partial s_2} , \label{eq:a0a1}
\end{eqnarray}  
where $s_p = \frac{1}{N} \sum_i (n_i)^p$. Note that, due to the re-summation of expansion (\ref{eq:explicit1}) of all terms up to order $N$, 
the present approach could not be anymore regarded as a $1/N$ correction.  
Energies obtained with this approximation are shown in figure \ref{fig3:pbcs} for various particle numbers and couplings. The energy found by minimizing the functional is overall in very good agreement with the exact energy for any particle number and coupling. 
\begin{figure}[htbq]
\includegraphics[width = 8cm]{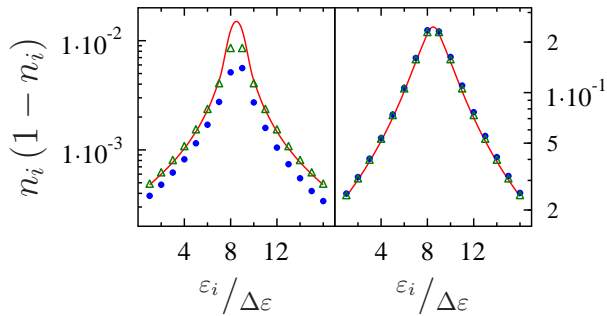}
\caption{ \label{fig5:pbcs} (Color online) Evolution of  $n_i (1-n_i)$ 
as a function of $\varepsilon_i/\Delta \varepsilon$  for the exact (solid line), PBCS (open triangles), and the new functional (filled circles) for $A=16$
and $g/\Delta \varepsilon=0.16$ (left) and $0.56$ (right).}
\end{figure}
A careful analysis shows a slight underestimation of the energy in the intermediate 
coupling regime associated also with a slight difference in the occupation numbers (see figure \ref{fig5:pbcs}).
This small discrepancy stems from the linear approximation made for the $\alpha_i$. 
%This is clearly seen when comparing the dependence of these coefficients as a function of $n_i$
%in the PBCS case and the present functional both taken at the minimum of the corresponding 
%energy (figure \ref{fig4:pbcs}).  
%\begin{figure}[htbq]
%\includegraphics[width = 8cm]{fig4pbcs.eps}
%\caption{ \label{fig4:pbcs} (Color online) Evolution of the coefficients $\alpha_i$ as a function of $n_i$ (left)
%or $\varepsilon_i$ (right)  at the minimum of energy.
%The different curves correspond to the PBCS result for $g/\Delta \varepsilon=0.32$ (dashed line), $0.64$ 
%(dotted line) and $0.96$ (solid line). The corresponding results obtained with the linear approximation (Eq. 
%(\ref{eq:linear})) are displayed by  filled circles, crosses and  open circles respectively. }
%\end{figure}
%While the BCS would give $\alpha_i =1$ for any particle 
%number and coupling, the new functional follows closely the PBCS evolution especially when coupling increases. The dependency 
%of $\alpha_i$ obtained in the PBCS case for small coupling also shows that a simple 
%linear approximation cannot fully grasp the physics of correlation in that case. 
To better account for occupation number behavior, 
%Following the same strategy as 
%above, 
quadratic or cubic corrections to (\ref{eq:linear}) might eventually be obtained. However, this will add 
complexity to the functional while the energy is already rather well reproduced. 
  
In this letter,  a new approach is proposed to account for pairing in finite systems using functional theories.
To escape difficulties recently observed \cite{Dob07,Lac09,Ben09,Dug09}, namely divergences and jumps,
the introduction of an auxiliary QP state
is avoided and a Projected BCS trial state is directly used. 
%The energy is then directly written as a functional of occupation numbers and natural states of the projected state. The new functional is benchmarked here on a schematic 
%pairing Hamiltonian, for which an exact solution can be obtained numerically.
In the pairing hamiltonian, a perfect agreement between the functional result and the exact 
energy is obtained at all coupling strength and particle number. The present method can be directly 
implemented on existing functional theories and should provide an accurate way to treat finite size 
effects. Note that present approach can easily be extended to odd systems \cite{Lac10}
and might  provide a tool of choice to study dynamics and thermodynamics of finite systems with pairing.  
Guided by the BCS theory, system at  finite temperature can be studied minimizing the free energy defined 
through
\footnote{since all the information on the system is now contained in the single-particle components, the entropy 
simply identifies with ${\cal S} [n_i]=- \sum_i [ n_i \log(n_i) + 
(1-n_i) \log(1 - n_i) ]$ \cite{Bal99}.}:
\begin{eqnarray}
{\cal F} [\{n_i\}, \varphi_i]  = {\cal E}' [\{n_i\}, \varphi_i] - T {\cal S} [n_i] .
\end{eqnarray}Note that, the present theory has been recently applied to nuclei showing that it solves recent difficulties 
associated to broken symmetries \cite{Dob07,Lac09} opening new perspectives.

%we show that the pairing Hamiltonian can be  accurately described by a functional of occupation numbers 
%in the spirit of density matrix functional theory. Since the functional is based on the PBCS ansatz, it is 
%expected to significantly improve the BCS functional and to be accurate within the range of application of
%PBCS.  Although the application presented here correspond to a rather schematic
%Hamiltonian, since only general properties of the projected state have been used  to get the functional, it is also expected to
%apply 
%for general Hamiltonian. 
%This functional approach to pairing will be particularly useful for systems with limited number of particles like ultrasmall metallic grains or nuclear systems. Here, we focus on ground state properties where 
%the recurrence relation method or direct projection can already be used to perform
%variation after projection \cite{San08,Von01}. These methods can however hardly be extended for the study 
%of finite temperature or time-dependent properties of systems with pairing.  The present functional 
%is anticipated to provide a tool of choice for these processes.
 
\begin{acknowledgments}  
 We are particularly grateful to N. Sandulescu for providing us with the exact Richardson and PBCS 
codes as well as helpful comments. We also thank Th. Duguet for helpful discussions. We also thank K. Washiyama and P. Mei for proofreading the manuscript.
\end{acknowledgments}

\end{document}